\documentclass[aps,prd,onecolumn,groupedaddress,showpacs,nofootinbib,amssymb]{revtex4}
\usepackage[dvips]{graphicx}
\usepackage{amssymb}
\usepackage{amsmath}
\usepackage{graphicx,,color}
\usepackage{amsfonts}
\usepackage{bm}
\usepackage{cancel}

\allowdisplaybreaks[4]

\begin{document}

\tolerance=5000

\newcommand\be{\begin{equation}}
\newcommand\ee{\end{equation}}
\newcommand\nn{\nonumber \\}
\newcommand\e{\mathrm{e}}

\title{Covariant Generalized Holographic Dark Energy
and Accelerating Universe}

\author{
Shin'ichi Nojiri$^{1, 2}$\footnote{E-mail address:
nojiri@phys.nagoya-u.ac.jp}
and
S.~D.~Odintsov$^{3,4}$\footnote{E-mail address:
odintsov@ieec.uab.es}
}

\affiliation{
$^1$Department of Physics, Nagoya University, Nagoya 464-8602, Japan \\
$^2$Kobayashi-Maskawa Institute for the Origin of Particles and the Universe,
Nagoya University, Nagoya 464-8602, Japan \\
$^3$ICREA, Passeig Luis Companys, 23, 08010 Barcelona, Spain \\
$^4$Institute of Space Sciences (IEEC-CSIC) C. Can Magrans s/n,
08193 Barcelona, Spain
}

\begin{abstract}

We proposed the generalized holographic dark energy model where infrared
cutoff is identified with the combination of the FRW universe parameters:
the Hubble rate, particle and future horizons, cosmological constant,
the universe life-time (if finite) and their derivatives.
It is demonstrated that with the corresponding choice of the cutoff one can
map such holographic dark energy to modified gravity or gravity with general
fluid.
Explicitly, $F(R)$ gravity and general perfect fluid are worked out in detail and
corresponding infrared cutoff is found.
Using this correspondence,  we get realistic inflation or viable dark energy or
unified inflationary-dark energy universe in terms of covariant holographic
dark energy.

\end{abstract}


\maketitle

\section{Introduction \label{Sec1}}

Quantum field considerations may play the fundamental role in the study of the
early- and late-universe evolution.
Indeed, if we consider a particle with mass $m$, the quantum correction to
the (flat-space) vacuum energy density $\rho_\mathrm{vacuum}$ is given by
\be
\label{q1}
\rho_\mathrm{vacuum} = \pm \frac{1}{\left( 2 \pi \right)^3}\int d^3 k
\left( \frac{1}{2} \sqrt{ k^2 + m^2 } \right) \, .
\ee
Here $+$ ($-$) sign corresponds to the bosonic (fermionic) particle.
For large $k$, the integration diverges but because the observed value of the
vacuum energy is very small, that is, $\left( 10^{-3}\, \mathrm{eV} \right)^4$,
the divergence could be absorbed or cancelled by some yet unknown
mechanism.\footnote{ In order to solve this problem, a topological model has
been proposed \cite{Nojiri:2016mlb} and the cosmology in this model has been
investigated \cite{Saitou:2017zyo,Mori:2017dhe}
}
On the other hand, if we conisder small $k$, which corresponds to the large
distance, there should be a minimum for $k$, nor there is, at least, a problem in
the causality.
We call the minimum as an infrared cutoff and denote it by
$\Lambda_\mathrm{inf}$.
Then we find $k\gtrsim \Lambda_\mathrm{inf}$.
If the minimum is really related with the causality, the infrared cutoff
$\Lambda_\mathrm{inf}$ is related with the horizon radius $L_\mathrm{H}$ as
$\Lambda_\mathrm{inf}\sim 1/L_\mathrm{H}$.
Then the vacuum energy  (\ref{q1}) could be estimated as
$\rho_\mathrm{vacuum} \sim \mp m \Lambda_\mathrm{inf}^3$ if the particle
has non-vanishing mass, $m\neq 0$ or
$\rho_\mathrm{vacuum} \sim \mp \Lambda_\mathrm{inf}^4$ if the particle is
massless, $m=0$.
Hence, the small but non-vanishing vacuum energy might be regarded as the
universe dark energy and in principle, might generate the accelerating
expansion of the current universe.

Of course, the real situation could be much more complicated.
For example, the infrared cutoff might be given by the temperature of the
current universe $T\sim 10^{-3}\, \mathrm{eV}$ then the energy scale of the
vacuum energy in the present universe can be naturally given if we consider
the massless particle,
$\rho_\mathrm{vacuum} \sim \mp \Lambda_\mathrm{inf}^4 \sim T^4 \sim
\left( 10^{-3}\, \mathrm{eV}\right)^4$.
We should also note that the Stefan-Boltzmann law tells $\rho \propto T^4$
although only the thermal energy cannot generate the accelerating expansion.
In the curved space-time, the cutoff could depend on the space-time
curvature or we may use the Hubble horizon as the horizon radius
$L\sim 1/H$, with the Hubble rate $H$.

If we consider the holographic principle \cite{Li:2004rb} (see also
\cite{Huang:2004wt, Wang:2004nqa, Myung:2005sv, Myung:2005pw,
Enqvist:2004xv, Hsu:2004jt, Ito:2004qi, GonzalezDiaz:2004tb, Huang:2004mx,
Nobbenhuis:2004wn, Gong:2004cb, Medved:2005di, Zhang:2005yz, Gong:2005ya,
Padmanabhan:2005cw, Guberina:2005fb,Arevalo:2013tta,
Chimento:2012zz,Zhang:2012qra, Zhang:2005hs}), the vacuum energy could be
proportional to the   universe radius square $L$,
\be
\label{q2}
\rho_\mathrm{vacuum} =\frac{3c^2}{\kappa^2 L^2}\, .
\ee
Here $\kappa$ is the gravitational coupling and $c$ is a constant.
The problem of the original holographic dark energy model \cite{Li:2004rb}
where the infrared cutoff was chosen as the size of the event horizon is the
fact that the corresponding FRW equations often do not correspond to any
covariant gravity theory and even may not predict the universe acceleration.
Subsequently, the generalized holographic dark energy has been proposed in
Ref.~\cite{Nojiri:2005pu} where infrared cutoff is identified with combination of
the FRW universe parameters: the Hubble constant, particle and future
horizons, cosmological constant and universe life-time (if finite).
Implicitly, the dependence from
the derivatives of the corresponding FRW parameters was also assumed.
Different versions of the cutoff corresponding to generalized holographic dark
energy \cite{Nojiri:2005pu} have been considered in
]Refs.~\cite{Granda:2008dk,Khurshudyan:2016gmb,Wang:2016och,
Khurshudyan:2016uql,KhodamMohammadi:2012ww,Belkacemi:2011zk,
Zhang:2011zze,Setare:2010zy,Nozari:2009zk,Sheykhi:2009dz,
Xu:2009xi,Wei:2009au,Setare:2008hm,Saridakis:2007wx,Setare:2006yj,
Felegary:2016znh}.

In this paper, we consider generalized holographic dark energy with arbitrary
cutoff  which depends on FRW universe parameters for the vacuum energy
density $\rho_\mathrm{vacuum}$.
In fact, the choice of the infrared cutoff could be included in the definition of
the quantum theory in curved spacetime.
Therefore, the unitarity and causality could give the important hints to find the
correct definition of the infrared cutoff.
The AdS/CFT might also give some indication because the infrared cutoff
could be related with the ultraviolet cutoff in the corresponding gauge theory.
However, there is no any definite prediction for the choice of the infrared
cutoff at least at present.
Then, we consider several possibilities for cutoff choice.
In the next section, we briefly review the holographic dark energy and
introduce the covariant generalized holographic dark energy model.
Section~\ref{Sec3} is devoted to generalized holographic dark energy which is
equivalent to $F(R)$ gravity. In this way, the consistent inflation naturally
emerges from covariant generalized holographic dark energy.
In the last section, the reconstruction of the arbitrary fluid as generalized
holographic dark energy is developed. The occurence of dark energy universe
or even unified inflation-dark energy universe in terms of such theory is
demonstrated.

\section{Generalized Holographic Dark Energy \label{Sec2}}

Let us consider the spatially-flat FRW Universe
\be
\label{JGRG14}
ds^2 = - dt^2 + a(t)^2 \sum_{i=1,2,3} \left(dx^i\right)^2\, .
\ee
Defining the Hubble rate $H=\dot a(t)/a(t)$, the first FRW equation has the
following form
\be
\label{FRW1}
\frac{3}{\kappa^2}H^2=\rho_\Lambda\, ,
\ee
which gives
\be
\label{H2} H=\frac{c}{L}\, .
\ee
Here we assume that $c$ is positive constant because  the expanding
universe
is considered.
It is known that if one chooses the infrared cutoff $\Lambda_\mathrm{inf}$ to
be the Hubble rate $H$, the accelerating universe cannot be realized.
There are more possibilities for the choice of infrared radius $L$.
For example, one may choose $L$ to be the particle horizon $L_p$ or the
future horizon $L_f$, which is defined by
\be
\label{H3}
L_p (t) \equiv a(t) \int_0^t\frac{dt'}{a(t')}\, ,\quad L_f(t)\equiv a(t)
\int_t^\infty \frac{dt'}{a(t')}\, .
\ee
For the FRW metric with the flat spatial part in (\ref{JGRG14}), by choosing
$L$ as $L_p$ or $L_f$, we find the following equation,
\be
\label{H5}
\frac{d}{dt}\left(\frac{c}{a(t) H(t)}\right)=\pm \frac{1}{a(t)}\, .
\ee
Here, the $+$ ($-$) sign corresponds to the particle (future) horizon.
We can easily solve Eq.~(\ref{H5}) and find
\be
\label{H6}
a(t) =a_0 t^{h_0}\, ,
\ee
with
\be
\label{H7}
h_0=\frac{1}{1\pm \frac{1}{c}}\, .
\ee
Then, in case $L=L_f$, the universe is accelerating because $h_0>1$.
When $c>1$ in case $L=L_p$, $h_0$
becomes negative and the universe is shrinking.
If one can change the direction of time as $t\to t_s -t$, instead of (\ref{H6}),
we find
\be
\label{H8}
a(t) =a_0\left(t_s - t\right)^{h_0}\, .
\ee
Then there will be a Big Rip singularity at $t=t_s$.
Because we change the direction of time, the particle horizon becomes a
future-like one,
\be
\label{H8b}
L_p(t) \to \tilde L_f(t) \equiv a(t) \int_t^{t_s}\frac{dt'}{a(t')}
= a(t) \int_{a(t)}^\infty \frac{da}{Ha^2}\, .
\ee
Note that if we choose $L$ as a future horizon $L=L_f$, there is a solution
describing the de Sitter space-time
\be
\label{dS1}
a(t) =a_0\e^{\frac{t}{l}}\quad \left(H=\frac{1}{l}\right) \, .
\ee
If we choose, however the particle horizon as $L$, there does not exist the
solution describing the  de Sitter space-time.
Additionally to the fact that not all choices of cutoff may lead to the
accelerating universe, it is easy to see that corresponding FRW equations
cannot be obtained from some covariant action.

In general, $L_\Lambda$ could be a combination (a function) of both, $L_p$,
$L_f$ \cite{Elizalde:2005ju}.
Furthermore, if there is a Big Rip singularity at $t=t_s$ and therefore the
lifetime of the universe is finite, $L$ can be also a function of $t_s$.
More general, there could be a case that $L$ depends on the Hubble rate $H$
and also the curvature as we have mentioned (see also
Ref.~\cite{Nojiri:2005pu}),
\be
\label{hp6}
L=L\left(L_p, \dot L_p, \ddot L_p, \cdots, L_f, \dot L_f, \ddot L_f, \cdots, t_s,
H, \dot H, \ddot H, \cdots \right)\, .
\ee
Some of the cutoffs given by (\ref{hp6}) cannot be obtained from the
covariant gravity theory.
Still such a possibility should not be excluded because the FRW background
breaks the covariance or at least the Lorentz symmetry, in some sense,
spontaneously.
A similar example might be the Casimir force, which could also appear
because we break the Lorentz invariance by the boundary conditions.
We call the the theory with above cutoff as covariant generalized holographic
dark energy in the case where it may be equivalently described by some
covariant theory.
Later on, we give some examples of such theory where the equivalence with
modified gravity or fluid theory is established.

Let us demonstrate that the above model may  unify the early-time inflation
and the current accelerating expansion of the universe.
We consider the case that $c=1$ and $L$ is given by
\be
\label{hlgr1}
\frac{1}{L} = \left( 1 - \frac{1}{t_0 h_0} \right) \frac{1}{L_f}
+ \frac{2h_0}{t_0 h_1} + \frac{h_1}{t_0}
\left( 1 - \frac{h_0^2}{h_1^2} \right) L _f \, .
\ee
Here $L_f$ is a future horizon defined by (\ref{H3}) and $t_0$, $h_0$, and
$h_1$ are positive constants and it is assumed $h_0>h_1$.
Then the solution of the first FRW equation is given by
\be
\label{hlgr}
H = h_0 - h_1 \tanh \frac{t}{t_0}
+ \frac{h_1}{t_0 \left( h_0 - h_1 \tanh \frac{t}{t_0} \right) \cosh^2
\frac{t}{t_0} }\, .
\ee
In fact, one can check that
\be
\label{hlgr2}
\frac{1}{L_f} = h_0 - h_1 \tanh \frac{t}{t_0} \, ,
\ee
Then at the early universe $t\to - \infty$, the Hubble rate $H$ goes to a
constant $H\to h_0 + h_1$, which may be identified with the inflation and at
the late universe $t\to +\infty$, $H$ goes to a constant again,
$H\to h_0 - h_1$, which may be identified with the late-time accelerating
expansion.
Instead of (\ref{hlgr1}), by using the scalar curvature
$R=6 \left( 2 H^2 + \dot H \right)$, we may discuss the model,
\be
\label{hlgr3}
\frac{1}{L^2} = \frac{R}{12} - \frac{1}{2} \left\{ 1 - \frac{1}{h_1^2}
\left( h_0 - \frac{1}{L} \right)^2 \right\} \left\{ \frac{h_1^2}{t_0^2}
\left( 1 - \frac{h_0^2}{h_1^2} \right) L_f^2 - \frac{h_1}{t_0}
+ \frac{1}{t_0^2} \right\} \, .
\ee
Even in the model (\ref{hlgr3}), the Hubble rate $H$ (\ref{hlgr}) is again the
solution.
Hence, generalized holographic dark energy may unify the inflation with dark
energy.

\section{Holographic Description of $F(R)$ gravity \label{Sec3}}

Let us now demonstrate the correspondence between the generalized
holographic dark energy of previous section and $F(R)$ gravity with the
action:
\be
\label{FR}
S = \frac{1}{2\kappa^2} \int d^4x \sqrt{-g} F(R) = \frac{1}{2\kappa^2}
\int d^4x \sqrt{-g} \left( R + f(R) \right) \, .
\ee
Because
\be
\label{hlgr4}
\dot L_p = H L_p + 1\, , \quad \dot L_f = H L_f - 1\, ,
\ee
we find
\be
\label{hlgr5}
H = \frac{\dot L_p}{L_p} - \frac{1}{L_p} = \frac{\dot L_f}{L_f} + \frac{1}{L_f}\, ,
\quad \dot H = \frac{\ddot L_p}{L_p} - \frac{{\dot L_p}^2}{L_p^2}
+ \frac{\dot L_p}{L_p^2}
= \frac{\ddot L_f}{L_f} - \frac{{\dot L_f}^2}{L_f^2}
  - \frac{\dot L_f}{L_f^2}\, ,
\ee
and therefore
\begin{align}
\label{hlgr6}
R_{tt}=& -3\left(\dot H + H^2\right)
= -3\left( \frac{\ddot L_p}{L_p} - \frac{\dot L_p}{L_p^2} + \frac{1}{L_p^2} \right)
= -3\left( \frac{\ddot L_f}{L_f} + \frac{\dot L_f}{L_f^2} + \frac{1}{L_f^2} \right)
\, , \nn
R_{ij}=& a^2 \left(\dot H + 3H^2\right)\tilde g_{ij}
=  a^2 \left( \frac{\ddot L_p}{L_p} + \frac{2{\dot L_p}^2}{L_p^2}
  - \frac{5 \dot L_p}{L_p^2} + \frac{3}{L_p^2} \right)\tilde g_{ij}
= a^2 \left(\frac{\ddot L_f}{L_f} + \frac{2{\dot L_f}^2}{L_f^2}
+ \frac{5 \dot L_f}{L_f^2} + \frac{3}{L_f^2} \right)\tilde g_{ij} \, ,\nn
R=& 6\dot H + 12 H^2 = R_L
\equiv 6 \left( \frac{\ddot L_p}{L_p} + \frac{{\dot L_p}^2}{L_p^2}
  - \frac{3\dot L_p}{L_p^2} + \frac{2}{L_p^2} \right)
= 6 \left( \frac{\ddot L_f}{L_f} + \frac{{\dot L_f}^2}{L_f^2}
+ \frac{3\dot L_f}{L_f^2} + \frac{2}{L_f^2} \right) \, .
\end{align}
In the case of $F(R)$ gravity (for general review, see
\cite{Nojiri:2010wj,Capozziello:2011et}) we have the following FRW equations
\begin{align}
\label{JGRG15L}
3H^2  =& -\frac{f(R)}{2} + 3\left(H^2 + \dot H\right) f'(R)
  - 3 H \frac{d f'(R)}{dt} + \kappa^2 \rho_\mathrm{matter}\, ,\nn
  - 3 H^2 - 2 \dot H =& \frac{f(R)}{2} - \left(\dot H + 3H^2\right) f'(R)
+ 6H \frac{d f'(R)}{dt} + \frac{d^2 f'(R)}{dt^2} + \kappa^2 p_\mathrm{matter}\, .
\end{align}
Using the holographic language,
we may rewrite (\ref{JGRG15L}) as follows
\begin{align}
\label{JGRG15L2}
3H^2  = & -\frac{f(R_L)}{2}
+ 3\left( \frac{\ddot L_p}{L_p} - \frac{\dot L_p}{L_p^2}
+ \frac{1}{L_p^2} \right)
f'(R_L) - 3 \left( \frac{\dot L_p}{L_p} - \frac{1}{L_p} \right) \frac{d f'(R_L)}{dt}
+ \kappa^2 \rho_\mathrm{matter}\, ,\nn
= & -\frac{f(R_L)}{2}
+ 3\left( \frac{\ddot L_f}{L_f} + \frac{\dot L_f}{L_f^2}
+ \frac{1}{L_f^2} \right)
f'(R_L) - 3 \left( \frac{\dot L_f}{L_f} + \frac{1}{L_f} \right) \frac{d f'(R_L)}{dt}
+ \kappa^2 \rho_\mathrm{matter}\, , \nn
  - 3 H^2 - 2 \dot H =& \frac{f(R_L)}{2}
  - \left( \frac{\ddot L_p}{L_p} + \frac{2{\dot L_p}^2}{L_p^2}
  - \frac{5 \dot L_p}{L_p^2} + \frac{3}{L_p^2} \right)f'(R_L)
+ 6\left( \frac{\dot L_p}{L_p} - \frac{1}{L_p} \right)
\frac{d f'(R_L)}{dt}
+ \frac{d^2 f'(R_L)}{dt^2} \nn
& + \kappa^2 p_\mathrm{matter} \nn
=& \frac{f(R_L)}{2}
  - \left(\frac{\ddot L_f}{L_f} + \frac{2{\dot L_f}^2}{L_f^2}
+ \frac{5 \dot L_f}{L_f^2} + \frac{3}{L_f^2} \right) f'(R_L)
+ 6 \left( \frac{\dot L_f}{L_f} + \frac{1}{L_f} \right)
\frac{d f'(R_L)}{dt} + \frac{d^2 f'(R_L)}{dt^2} \nn
& + \kappa^2 p_\mathrm{matter}\, .
\end{align}
The above correspondence clearly shows how the arbitrary $F(R)$ gravity
may be mapped into the covariant generalized holographic dark energy.
Similarly, the equivalence with other modified gravities like modified
Gauss-Bonnet gravity, non-local gravity, string-inspired theory may be
established.

As an example, we consider the case that
\be
\label{Rsq1}
f(R) = \alpha R^2\, ,
\ee
with a constant $\alpha$.
This Starobinsky inflation model gives the following spectral index $n_s$ and
the scalar-tensor ratio $r$ \cite{Hinshaw:2012aka},
\be
\label{nsrR2}
n_s = 0.967\, , \quad r = 3.33 \times 10^{-3}\, ,
\ee
which is consistent with results in Planck 2015 \cite{Array:2015xqh},
\be
\label{nsrR2B}
n_s = 0.968\pm 0.006\, (68\% \mathrm{CL})\, , \quad
r < 0.11\, (95\% \mathrm{CL}) \, .
\ee
For the model (\ref{Rsq1}), one gets
\be
\label{Rsq2}
\frac{c^2}{\kappa^2 L^2}
= 18 \alpha \left( - \frac{2 \dddot L_p \dot L_p}{L_p^2}
+ \frac{2 \dddot L_p}{L_p^2} - \frac{2 \ddot L_p {\dot L_p}^2}{L_p^3}
+ \frac{{\ddot L_p}^2}{L_p^2} + \frac{4 \ddot L_p \dot L_p}{L_p^3}
  - \frac{8 \ddot L_p}{L_p^3} + \frac{3{\dot L_p}^4}{L_p^4}
  - \frac{14 {\dot L_p}^3}{L_p^4} + \frac{21 {\dot L_p}^2}{L_p^4}
  - \frac{10 \dot L_p}{L_p^4} \right)\, .
\ee
Hence, the correspondence between the $F(R)$ gravity  and the covariant
generalized holographic dark energy models is established.
This equivalence shows how to describe the accelerating universe in terms of
generalized holographic dark energy.

\section{Generalized Holographic Dark Energy Description of Perfect Fluid
\label{Sec4}}

One may further develop the correspondence between the holographic dark
energy and the perfect fluid.
The equation of state (EoS) of the general perfect fluid is given by
\be
\label{EoS1}
p = - \rho + h(\rho) \, .
\ee
Here $p$ is the pressure and $\rho$ is the energy density and $h(\rho)$ is a
function of the energy density $\rho$.
By using (\ref{hlgr5}), the conservation law
\be
\label{EoS2}
0 = \dot\rho + 3 H \left( \rho + p \right) = \dot\rho + 3 H h(\rho) \, ,
\ee
can be rewritten as follows,
\be
\label{EoS3}
0 = \dot\rho + 3 \left( \frac{\dot L_p}{L_p} - \frac{1}{L_p} \right) h(\rho)
= \dot\rho + 3 \left( \frac{\dot L_f}{L_f} + \frac{1}{L_f} \right) h(\rho)\, ,
\ee
which gives
\be
\label{EoS4}
\ln \frac{L_p}{L_0} - \int^t \frac{dt}{L_p} = \ln \frac{L_f}{L_0}
+ \int^t \frac{dt}{L_f} = \frac{1}{3} \int \frac{d\rho}{h(\rho)}\, .
\ee
Here $L_0$ is a constant which is introduced due to a dimensional
consideration.
Eq.~(\ref{EoS4}) can be algebraically solved with respect to $\rho$,
\be
\label{EoS5}
\rho = \rho \left( \ln \frac{L_p}{L_0} - \int^t \frac{dt}{L_p} \right)
= \rho \left( \ln \frac{L_f}{L_0} + \int^t \frac{dt}{L_f} \right) \, .
\ee
Then we find that the perfect fluid is also described by the holographic
language.

As the first simple example, we may consider the fluid with constant EoS
parameter $w$, $p=w\rho$, that is
\be
\label{EoS6}
h(\rho) = \left( 1 + w \right) \rho\, ,
\ee
which gives $a(t) \propto t^{\frac{2}{3(w+1)}}$.
Then Eq.~(\ref{EoS4}) gives
\be
\label{EoS7}
\ln \frac{L_p}{L_0} - \int^t \frac{dt}{L_p} = \ln \frac{L_f}{L_0}
+ \int^t \frac{dt}{L_f} = \frac{1}{3 \left( 1 + w \right)}
\ln \frac{\rho}{\rho_0} \, ,
\ee
that is,
\be
\label{EoS8}
\rho = \rho_0 \left( \frac{L_p}{L_0}\right)^{3 \left( 1 + w \right)}
\e^{- 3 \left( 1 + w \right) \int^t \frac{dt}{L_p}}
= \rho_0 \left( \frac{L_f}{L_0}\right)^{3 \left( 1 + w \right)}
\e^{3 \left( 1 + w \right) \int^t \frac{dt}{L_f}} \, .
\ee
For general perfect fluid, by using the FRW equations
\be
\label{EoS9}
\rho = \frac{3}{\kappa^2}H^2\, ,\quad p
=-\frac{1}{\kappa^2}\left(2\dot H + 3H^2\right) \, .
\ee
one can reconstruct the holographic cutoff.
If the Hubble rate $H$ is given as a function of the cosmological time $t$,
$H=J(t)$,
\be
\label{EoS10} p=-\rho - \frac{2}{\kappa^2}J'\left(J^{-1}
\left(\kappa\sqrt{\frac{\rho}{3}}\right)\right) \, ,
\ee
Eq.~(\ref{EoS10}) can be regarded as a generalized equation of state, where
\be
\label{EoS11}
h(\rho) = - \frac{2}{\kappa^2}J'\left(J^{-1}
\left(\kappa\sqrt{\frac{\rho}{3}}\right)\right) \, .
\ee
In general, it is difficult to write down the explicit form of $h(\rho)$ for general
$J(t)$ and also to solve (\ref{EoS4}) explicitly.
We now just give an example, where
\be
\label{EoS12}
J(t) = J_1 t + J_2 t^{-1}\, .
\ee
By using (\ref{EoS10}), we may define the effective EoS parameter
$w_\mathrm{eff}$ by
\be
\label{EoS13}
w_\mathrm{eff} = \frac{p}{\rho}= - 1 - \frac{2 \dot H}{3H^2}\, .
\ee
Then $H=J(t)$ in (\ref{EoS12}) gives
\be
\label{EoS14}
w_\mathrm{eff} = - 1 - \frac{2 \left( J_1 - J_2 t^{-2} \right)}{ 3\left( J_1 t
+ J_2 t^{-1} \right)^2} \, .
\ee
One finds that for $t\to +\infty$, $w_\mathrm{eff} \to -1$, which corresponds
to the accelerating expansion.
On the other hand, when $t\to 0$, $w_\mathrm{eff}$ behaves as
$w_\mathrm{eff} \to -1 + \frac{2}{3J_2}$.
Especially if $J_2=\frac{2}{3}$, the effective EoS parameter $w_\mathrm{eff}$
coincides with the dust EoS parameter.
Because Eq.~(\ref{EoS12}) can be solved with respect to $t$ as follows,
\be
\label{EoS15}
t = \frac{J \pm \sqrt{ J^2 - 4 J_1 J_2}}{2J_1}\, ,
\ee
Eq.~(\ref{EoS11}) gives
\be
\label{EoS16B}
h(\rho) = \frac{1}{\kappa^2 J_2} \left( \frac{\kappa^2 \rho}{3} \mp
\sqrt{ \frac{\kappa^2 \rho}{3} \left( \frac{\kappa^2 \rho}{3}
  - 4 J_1 J_2 \right)} \right)\, .
\ee
We should note the sign $\mp$ corresponds to the sign $\pm$ in
Eq.~(\ref{EoS15}).
Then Eq.~(\ref{EoS4}) gives
\begin{align}
\label{EoS16}
& \ln \frac{L_p}{L_0} - \int^t \frac{dt}{L_p} = \ln \frac{L_f}{L_0}
+ \int^t \frac{dt}{L_f} \nn &= \frac{1}{4J_1} \int \frac{d\rho}{\rho}
\left( \frac{\kappa^2 \rho}{3} \pm  \sqrt{ \frac{\kappa^2 \rho}{3}
\left( \frac{\kappa^2 \rho}{3} - 4 J_1 J_2 \right)} \right) \, .
\end{align}
It is a little bit tedious to execute the integration and rather difficult to solve
Eq.~(\ref{EoS16}) with respect to $\rho$ analytically.

In Ref.~\cite{Bamba:2014wda}, the following EoS was investigated,
\be
\label{EoS17}
p = - \left( 1 + \frac{\beta}{3} \right) \rho
+ \frac{G_3 \beta}{\kappa^2} \, ,
\ee
with constants $\beta$ and $G_3$.
The EoS fluid (\ref{EoS17}) gives the following Hubble rate,
\be
\label{EoS18}
H^2 = G_2 a^\beta + G_3\, .
\ee
With the EoS (\ref{EoS17}), $h(\rho)$ in (\ref{EoS1}) is given by
\be
\label{EoS19}
h(\rho) = - \frac{\beta}{3} \rho + \frac{G_3 \beta}{\kappa^2}\, .
\ee
Then Eq.~(\ref{EoS4}) gives
\be
\label{EoS20}
\ln \frac{L_p}{L_0} - \int^t \frac{dt}{L_p} = \ln \frac{L_f}{L_0}
+ \int^t \frac{dt}{L_f} = - \frac{1}{\beta} \int
\frac{d\rho}{\rho - \frac{3G_3 \beta}{\beta \kappa^2}}
= - \frac{1}{\beta} \ln \left( \frac{\rho
  - \frac{3G_3 \beta}{\beta \kappa^2}} {\rho_0} \right) \, ,
\ee
and we obtain
\be
\label{EoS21}
\rho = \frac{3G_3 \beta}{\beta \kappa^2}
+ \rho_0 \left(\frac{L_p}{L_0}\right)^{- \beta}
\e^{\beta \int^t + \frac{dt}{L_p} }
= \frac{3G_3 \beta}{\beta \kappa^2}
+ \rho_0 \left(\frac{L_f}{L_0}\right)^{- \beta}
\e^{-\beta \int^t \frac{dt}{L_f} } \, .
\ee
This shows the equivalence of specific perfect fluid universe with covariant
generalized holographic dark energy for specific cutoff.

Furthermore, in Ref.~\cite{Nojiri:2015wsa}, the following EoS was proposed:
\be
\label{C4}
p = - \rho - 2 \cdot 3^{- \frac{\alpha - 1}{2\alpha}} \alpha
\kappa^{- \frac{\alpha + 1}{\alpha}} f_0^{2 - \alpha}
\rho ^\frac{\alpha - 1}{2\alpha}\, .
\ee
Here $\alpha$ and $f_0$ are constants.
The universe inspired by the EoS fluid (\ref{C4}) evolves with the following
Hubble rate,
\be
\label{IV1B00}
H(t) = f_0 \left| t_s - t \right|^\alpha \quad \mbox{or} \quad
a(t) \propto \e^{ \frac{f_0}{\alpha+1} f_0 \left| t_s - t \right|^{\alpha + 1}}\, .
\ee
The form of the Hubble rate (\ref{IV1B00}) shows (see \cite{Nojiri:2005sx}):
\begin{itemize}
\item If $-1<\alpha<-\frac{1}{2}$, a Type III singularity occurs.
\item If $\alpha<-1$, a Type I singularity appears.
\item If $\alpha>1$, a Type IV singularity occurs.
\item If $-\frac{1}{2}<\alpha<1$, a Type II singularity occurs.
\end{itemize}
As the EoS fluid (\ref{C4}) gives
\be
\label{EoS22B}
h(\rho) = - 2 \cdot 3^{- \frac{\alpha - 1}{2\alpha}} \alpha
\kappa^{-\frac{\alpha + 1}{\alpha}} f_0^{2 - \alpha}
\rho^\frac{\alpha - 1}{2\alpha}\, ,
\ee
Eq.~(\ref{EoS4}) leads to
\be
\label{EoS20B}
\ln \frac{L_p}{L_0} - \int^t \frac{dt}{L_p} = \ln \frac{L_f}{L_0}
+ \int^t \frac{dt}{L_f}
= - \frac{1}{2 \cdot 3^{\frac{\alpha + 1}{2\alpha}} \alpha
\kappa^{-\frac{\alpha + 1}{\alpha}} f_0^{2 - \alpha}}
\int \frac{d\rho}{\rho^\frac{\alpha - 1}{2\alpha}}
= - \frac{1}{3^{\frac{\alpha + 1}{2\alpha}} \left( \alpha +1 \right)
\kappa^{-\frac{\alpha + 1}{\alpha}} f_0^{2 - \alpha}}
\rho^{\frac{\alpha + 1}{2\alpha}}\, .
\ee
Then we find
\be
\label{EoS21B}
\rho = \left( - 3^{\frac{\alpha + 1}{2\alpha}} \left( \alpha +1 \right)
\kappa^{-\frac{\alpha + 1}{\alpha}} f_0^{2 - \alpha} \left(
\ln \frac{L_p}{L_0} - \int^t \frac{dt}{L_p} \right) \right)^{\frac{2\alpha}{\alpha
+ 1}}
= \left( - 3^{\frac{\alpha + 1}{2\alpha}} \left( \alpha +1 \right)
\kappa^{-\frac{\alpha + 1}{\alpha}} f_0^{2 - \alpha} \left( \ln \frac{L_f}{L_0}
+ \int^t \frac{dt}{L_f} \right) \right)^{\frac{2\alpha}{\alpha + 1}} \, .
\ee
Again, the holographic description of the above perfect fluid is established.

As an example which unifies the inflation  and the late-time accelerating
expansion, one may consider the following model
\be
\label{EoS22}
\rho(t) = \frac{\Lambda_e + \Lambda_l a(t)}{1+a(t)}\, .
\ee
At the early universe, where $a(t)\to 0$, we find that $\rho$ goes to a
constant, $\rho \to \Lambda_e$ and even at the late universe, where
$a\to \infty$, $\rho$ goes to a constant $\rho \to \Lambda_l$.
Hence, at the early universe and at the late universe, the asymptotically
de Sitter universe is realized.
The scale factor $a$ behaves as
$a(t) \propto \e^{t \sqrt{\frac{\kappa^2 \Lambda_e}{3}}}$ at the early universe
and $a(t) \propto \e^{t \sqrt{\frac{\kappa^2 \Lambda_l}{3}}}$ at the late
universe.
The conservation law (\ref{EoS2}) gives
\be
\label{EoS23}
h \left( \rho \left(t\right) \right)
= -\frac{\Lambda_e - \Lambda_l}{3 \left( 1 + a(t) \right)^2}\, .
\ee
As Eq.~(\ref{EoS22}) can be solved with respect to $a(t)$ as
\be
\label{EoS24}
a(t) = \frac{\Lambda_e  - \rho(t)}{\rho(t) - \Lambda_l}\, ,
\ee
one gets
\be
\label{EoS25}
h(\rho) = \frac{\left( \rho - \Lambda_l \right)^2}{3
\left( \Lambda_e - \Lambda_l \right)}\, .
\ee
Eq.~(\ref{EoS4}) gives
\be
\label{EoS26}
\ln \frac{L_p}{L_0} - \int^t \frac{dt}{L_p} = \ln \frac{L_f}{L_0}
+ \int^t \frac{dt}{L_f} = \left( \Lambda_e - \Lambda_l \right)
\int \frac{d\rho}{\left( \rho - \Lambda_l \right)^2}
= - \frac{\left( \Lambda_e - \Lambda_l \right)} { \rho - \Lambda_l } \, ,
\ee
which can be solved with respect to $\rho$ as follows,
\be
\label{EoS27}
\rho = \Lambda_l - \frac{\left( \Lambda_e - \Lambda_l \right)}
{\ln \frac{L_p}{L_0} - \int^t \frac{dt}{L_p}}
= \Lambda_l - \frac{\left( \Lambda_e - \Lambda_l \right)} {\ln \frac{L_f}{L_0}
+ \int^t \frac{dt}{L_f}}\, .
\ee
Thus, we  succeeded to establish the correspondence between general
perfect fluid with Einstein gravity and the covariant generalized holographic
dark energy model. Due to well-known equivalence between fluid description
and scalar-tensor description (for review, see
\cite{Bamba:2012cp}) one can extend the above correspondence to
scalar-tensor theory.

\section{Discussion \label{Sec5}}

In summary, we proposed the generalized holographic dark energy model
where infrared cutoff is identified with combination of the FRW parameters:
the Hubble rate, particle and future horizons, cosmological constant and
universe life-time and the derivatives of the corresponding parameters. It is
pointed out that for  simple and natural choices of  the infrared cutoff
motivated by the considerations of the unitarity and causality or AdS/CFT
related hints the emerging universe may be not accelerating one.
Furthermore, it often happens that such holographic model does not admit the
covariant description.
However, as we demonstrate
in this letter, with more complicated choice of the infrared cutoff one can map
the (covariant) generalized holographic dark energy to the modified gravity or
to General Relativity with quite general fluid.
Specifically, the examples of the arbitrary $F(R)$ gravity and perfect fluids are
worked  out in detail.
It is explicitly shown how to get the realistic inflationary universe, or viable
dark energy universe or even the unification of the inflation with dark energy
epoch in frames of specific covariant generalized holographic dark energy.

The results of this letter give the clear recipe on how to rewrite the arbitrary
modified gravity including the scalar-tensor theory or gravity theory with fluid
matter as the covariant holografic model.
It may  indicate that different descriptions of the universe evolution may have
the common origin related with yet not fully understood symmetry somehow
related with holography.


We have established that  $F(R)$ gravity can be rewritten in the 
holographic
language at the level of background equivalence.
Then it might be interesting if we consider the equivalence at the perturbation
level.
In the stanard formulation of the  holographic dark energy, the infrared
cutoff only depends on the time coordinate but the scalar curvature also
depends on the space coordinates.
Hence,  in Eq.(\ref{hlgr6}), $L_p$ or $L_f$ should also
depend on the space coordinates when we consider the perturbation
although $L_p$ and $L_f$ should be usually spatially constant in
the standard holographic dark energy.
This indicates that there is no the equivalence at the 
perturbation
level between the holographic dark energy and the $F(R)$ gravity.
We expect that the spectral index $n_s$ and
the scalar-tensor ratio $r$ are  the same in both models but 
cosmological perturbations theory may lead to different results.
Furthermore, the infrared cutoff can depend on the point where
we are considering the theory. For example, near the black hole, the 
infrared cutoff
should depend on the spatial coordinates in addition to the 
time-coordinate.
Such an infrared cutoff could also depend on the non-local quantities as
the position of the black hole or the local fluctuation of the expansion
and therefore it could be rather difficult to formulate the perturbation 
in such
a complicated case.


As another example of the dark energy models,
we may consider the viscous dark energy model
(see {Brevik:2017msy} and references therein)
where a bulk viscosity in the cosmic fluid generates the accelerating
expansion of the universe.
In the viscous dark energy model, 
the conservation law of the fluid
is modified by the bulk viscosity $\zeta$ as follows,
\begin{equation}
\label{viscos}
\dot \rho + 3 H \left( \rho + p \right) = 9 \zeta H^2 \, .
\end{equation}
Then  FRW equations are modified as follows,
\begin{equation}
\label{viscos2}
3 H^2 = \kappa^2 \rho \, , \quad
  -3 H^2 - 2 \dot H = \kappa^2 \left( p - 3 \zeta H \right) \, .
\end{equation}
By comparing the Eq.(\ref{viscos2}) with Eq.(\ref{JGRG15L}),
we may identify
\begin{align}
\label{JGRG15L}
\rho  =& \frac{1}{\kappa^2} \left\{ -\frac{f(R)}{2}
+ 3\left(H^2 + \dot H\right) f'(R)
  - 3 H \frac{d f'(R)}{dt} \right\}\, ,\nn
p =& \frac{1}{\kappa^2} \left\{ \frac{f(R)}{2} - \left(\dot H + 3H^2\right) f'(R)
+ 6H \frac{d f'(R)}{dt} + \frac{d^2 f'(R)}{dt^2} \right\} + 9 \zeta H^2 \, .
\end{align}
In the holographic view point, $H$ can be given
by the infrared cutoff as shown in (\ref{hlgr5}).
Therefore the conservation law of the fluid in (\ref{viscos}) can be further
rewritten as,
\begin{align}
\label{viscoshol}
& \dot \rho + 3 H \left( \rho + p \right)
= \dot \rho + 3 \left( \frac{\dot L_p}{L_p} - \frac{1}{L_p}\right)
\left( \rho + p \right)
= 9 \zeta \left( \frac{\dot L_p}{L_p} - \frac{1}{L_p}\right) ^2 \nn
&=\dot \rho + 3 \left( \frac{\dot L_f}{L_f} - \frac{1}{L_f}\right)
\left( \rho + p \right)
= 9 \zeta \left( \frac{\dot L_f}{L_f} - \frac{1}{L_f}\right) ^2 \, .
\end{align}
Further by using more genenal infrared cutoff in (\ref{hp6}), we may write
(\ref{viscos})  as follows
\begin{equation}
\label{viscoshol2}
\dot \rho + 3 H \left( \rho + p \right)
=  9 \zeta H\left(L, \dot L, \ddot L, \cdots, t_s, \cdots \right)^2 \, .
\end{equation}
This  indicates that   viscous dark 
energy maybe related with
the holographic dark energy or vice-versa.


It might be also interesting to consider the holographic dark energy
in the brane cosmology
\cite{Saridakis:2007wx,Saridakis:2007ns,Saridakis:2007cy}.
In $D$ dimensional space-time, the total energy in the region with
a radiuss $r$ does not exceed the mass of the maximum black hole in the
region. In terms of the Schwarzschild radius $r_s$, the mass $M$
is proportional to $r_s^{D-3}$, $M \propto r_s^{D-3}$, then the energy
density $\rho$ is restricted to be
\begin{equation}
\label{rho}
\rho \leq \frac{\left. M \right|_{r_s=r}}{V} \propto r^{-2} \, .
\end{equation}
Here $V$ is the volume of the region, which is proportional to $r^{D-1}$.
Then by choosing $r$ to be the infrared cutoff $L$, we find
$\rho \propto L^{-2}$.
Hence, if we consider the $D-1$ dimensional brane in the $D$ dimensional
bulk space-time, the energy density which is proportional to $L^{-2}$ is
induced, its behavior coincides with the behavior of the four dimensional
holographic dark energy in (\ref{q2}).
Therefore the $F(R)$ or any other modified gravity in the bulk and/or the 
brane can be rewritten
as the holographic dark energy.
This will be considered elsewhere.

\section*{Acknowledgements}

This work is supported (in part) by
MEXT KAKENHI Grant-in-Aid for Scientific Research on Innovative Areas
``Cosmic Acceleration''  (No. 15H05890) (SN) and by MINECO (Spain), project
FIS2013-44881, FIS2016-76363-P(SDO) and by CSIC I-LINK1019 Project
(SDO and SN).

\end{document}